\documentclass[sn-mathphys,Numbered]{sn-jnl}% Math and Physical Sciences Reference Style
\usepackage{graphicx}%
\usepackage{multirow}%
\usepackage{amsmath,amssymb,amsfonts}%
\usepackage{amsthm}%
\usepackage{mathrsfs}%
\usepackage[title]{appendix}%
\usepackage{xcolor}%
\usepackage{textcomp}%
\usepackage{manyfoot}%
\usepackage{booktabs}%
\usepackage{algorithm}%
\usepackage{algorithmicx}%
\usepackage{algpseudocode}%
\usepackage{listings}%
\usepackage{mathtools}
\usepackage{caption}%

\usepackage{color} % for the command \textcolor
\usepackage{soul} % for the command \hl

\newcommand{\reals}{{\mathbb{R}}}

\usepackage{fancyhdr}
\begin{document}

\title{Spreading Code Optimization for Low-Earth Orbit Satellites via
Mixed-Integer Convex Programming}

%%=============================================================%%
%% Prefix	-> \pfx{Dr}
%% GivenName	-> \fnm{Joergen W.}
%% Particle	-> \spfx{van der} -> surname prefix
%% FamilyName	-> \sur{Ploeg}
%% Suffix	-> \sfx{IV}
%% NatureName	-> \tanm{Poet Laureate} -> Title after name
%% Degrees	-> \dgr{MSc, PhD}
%% \author*[1,2]{\pfx{Dr} \fnm{Joergen W.} \spfx{van der} \sur{Ploeg} \sfx{IV} \tanm{Poet Laureate} 
%%                 \dgr{MSc, PhD}}\email{iauthor@gmail.com}
%%=============================================================%%

\author[1]{\fnm{Alan} \sur{Yang}}\email{yalan@stanford.edu}
\author[2]{\fnm{Tara} \sur{Mina}}\email{tymina@stanford.edu}
\author*[1,2]{\fnm{Grace} \sur{Gao}}\email{gracegao@stanford.edu}

\affil[1]{%
\orgdiv{Department of Electrical Engineering},
\orgname{Stanford University},
\orgaddress{\street{350 Jane Stanford Way}, 
\city{Stanford}, 
\postcode{94305}, 
\state{CA}, 
\country{USA}}}

\affil[2]{%
\orgdiv{Department of Aeronautics \& Astronautics},
\orgname{Stanford University},
\orgaddress{\street{496 Lomita Mall}, 
\city{Stanford}, 
\postcode{94305}, 
\state{CA}, 
\country{USA}}}

%%==================================%%
%% sample for unstructured abstract %%
%%==================================%%

\abstract{Optimizing the correlation properties of spreading codes is critical
for minimizing inter-channel interference in satellite navigation systems. By
improving the codes' correlation sidelobes, we can enhance navigation
performance while minimizing the required spreading code lengths. In the case of
low earth orbit (LEO) satellite navigation, shorter code lengths (on the order
of a hundred) are preferred due to their ability to achieve fast signal
acquisition. Additionally, the relatively high signal-to-noise ratio (SNR) in
LEO systems reduces the need for longer spreading codes to mitigate
inter-channel interference. In this work, we propose a two-stage block
coordinate descent (BCD) method which optimizes the codes' correlation
properties while enforcing the autocorrelation sidelobe zero (ACZ) property. In
each iteration of the BCD method, we solve a mixed-integer convex program (MICP)
over a block of 25 binary variables. Our method is applicable to spreading code
families of arbitrary sizes and lengths, and we demonstrate its effectiveness
for a problem with 66 length-127 codes and a problem with 130 length-257 codes.}


%%\pacs[JEL Classification]{D8, H51}

%%\pacs[MSC Classification]{35A01, 65L10, 65L12, 65L20, 65L70}

\maketitle
\section{Introduction}

In this work, we consider the design of navigation spreading codes for future
low-Earth orbit (LEO) satellite constellations. LEO constellations are typically
comprised of hundreds, or even thousands, of low-cost satellites
\cite{ReidWELCGOW2020}. Commercial examples include Iridium \cite{Iridium},
OneWeb \cite{OneWeb}, and the constellations proposed by Samsung \cite{Samsung},
SpaceX \cite{SpaceX}, and Xona \cite{xona2023septentrio}. Several of those
constellations are primarily designed to deliver global internet coverage, but
can also be leveraged for navigation \cite{ReidNWE2016}. In addition, the
European space agency (ESA) recently announced the LEO-PNT program, which will
test capabilities for navigation and timing using LEO satellites
\cite{Dennehy2023}. See the review \cite{ProlFSVPEI2022} {for a survey of recent
developments in LEO PNT}.

All code-division multiple access (CDMA) systems, including satellite navigation
systems, rely on spreading codes \cite{MisraE2012}. In CDMA, each satellite
modulates its signal with a unique and known \emph{spreading code}, which is
typically a binary sequence. The receiver then correlates the received signal
with replicas of each satellite's spreading code to identify the source of the
transmitted signal. Therefore, it is necessary for the spreading codes to have
low pairwise cross-correlation to minimize inter-channel interference.
Additionally, it is important for the codes to have low autocorrelation
sidelobes to minimize the effects of multi-path and self-interference, and to
ensure that the time of signal reception can be correctly determined.

LEO systems suffer less path loss than navigation systems in medium-Earth orbit
(MEO) such as GPS, improving signal strength 1000-fold, or by 30dbB
\cite{ReidWELCGOW2020}. Therefore, short code lengths (on the order of a
hundred) may be preferable since short code lengths correspond to fast signal
acquisition, and the relatively high signal-to-noise ratio (SNR) means that
long spreading codes are not needed for low inter-channel interference
\cite{BekhitEE2011}.
% The pattern of a large number of codes with short lengths
% provides unique challenges and opportunities for spreading code optimization.

In this work, we propose a two-stage block coordinate descent (BCD) method for
finding good spreading codes satisfying the autocorrelation sidelobe zero (ACZ)
property \cite{WallnerAHR2007,Winkel2011}. The ACZ property requires that the
shift-one autocorrelation of each code is minimal. Imposing the ACZ constraint
is useful for ensuring that the tracking performance is consistently good across
all the satellites in the constellation. \cite{Winkel2011,WallnerAHR2007}. The
first stage of our BCD method finds a feasible code family that satisfies the
ACZ property, and the second stage minimizes the sum of squared auto- and
cross-correlation sidelobes, without breaking the ACZ property.

Our method is based on the recently-proposed mixed-integer convex program (MICP)
approach to spreading code optimization \cite{ConfortiCZ2014,
YangMG2023a,YangMG2023b}, which has shown promising results for optimizing
spreading codes for MEO applications. 
% This approach is particularly well-suited for optimizing spreading codes for LEO
% applications, since the code lengths can be made relatively short, on the order
% of a hundred. For shorter code lengths, the MICP approach may be used to
% optimize over 
In that approach, the spreading code optimization problem is formulated as a
mixed-integer convex program (MICP). In each iteration of BCD, we solve the
optimization problem exactly over a subset of the binary variables, with the
others held fixed. The partial minimization problem is also an MICP, and may be
solved using an MICP solver such as Gurobi \cite{Gurobi}. In this work, we
handle the ACZ constraint by using the fact that it may be enforced using linear
inequality constraints \cite{YangMG2023b}.

Unlike the prior works \cite{YangMG2023a,YangMG2023b}, we focus on the LEO
setting, where the number of codes is relatively large. In this regime, the
choice of the subset of binary variables to optimize over in each iteration of
BCD is an important consideration. We propose a variable subset selection
strategy that can achieve a small per-iteration runtime cost by tuning the
number of codes from which the binary variables are selected in a given
iteration. For a fixed variable subset size, choosing the variables from a small
number of codes can make the subproblems more difficult to solve, and can lead
large per-iteration runtime cost. On the other hand, choosing the variables from
a large number of codes can make the subproblems easier to solve, but can still
lead to a large per-iteration runtime cost due to the computational overhead
from setting up the MICP subproblems.

The rest of the paper is organized as follows. We review related work in
Subsection \ref{ss-related-work}. In Section \ref{s-optimization}, we describe
the spreading code optimization problem and the MICP formulation of the problem.
In Section \ref{s-bcd}, we describe the proposed BCD. Section \ref{s-results}
presents numerical results, and Section \ref{s-conclusion} concludes.

\subsection{Related work}\label{ss-related-work}

The problem of designing spreading codes with good correlation properties has a
long history. Current satellite navigation systems, such as GPS, use
pseudo-random noise (PRN) codes such as Gold codes \cite{Gold1967}, which can be
generated using linear-feedback shift registers, and the Weil codes
\cite{Rushanan2007,Legendre1808}, which are based on Legendre sequences. While
those codes satisfy provable bounds on their auto- and cross-correlation, they
are only available for specific lengths and number of codes, and cannot be
easily modified. For example, truncating or extending those codes by even a
single bit can compromise their correlation properties \cite{WallnerAHR2007}.

To address those limitations, there has been growing interest in designing the
spreading codes by directly optimizing the auto- and cross-correlation.
Population-based methods, such as genetic algorithms \cite{MinaG2019,LiuSW2022},
natural evolution strategies \cite{MinaG2022}, and the cross-entropy method
\cite{MinaYG2023} have been applied, and the European Union's Galileo
constellation uses spreading codes designed by a genetic algorithm
\cite{WallnerAHR2007, WallnerAWHI2008}. However, those methods do not consider
the structure in the objective, and often require extensive tuning in order to
work well. In addition, they have focused on codes for medium-Earth orbit (MEO)
constellations such as GPS \cite{MinaG2022,YangMG2023b} and Galileo
constellations \cite{WallnerAHR2007,WallnerAWHI2008}. In those settings, the
code length is orders of magnitude larger than the number of the codes; this is
neccessary for good system performance due to the relatively lower SNR in the
MEO setting. 

Coordinate descent and BCD methods have been proposed to optimize sets of binary
sequences for multiple-input multiple-output (MIMO) radar systems
\cite{AlaeeMN2019, CuiYFHL2017, LinLi2020, HuangL2020,YangMG2023a}. Like the
spreading codes for navigation systems, the binary sequences used in those
applications are required to have low auto- and cross-correlation. However, the
aforementioned works only optimize over either a single binary variable at a
time, or a small number (\emph{e.g.}, four), using an exhaustive search. In
contrast, by using branch and bound to perform the BCD updates
\cite{YuanLZ2017,YangMG2023a,YangMG2023b}, we can efficiently optimize over
large blocks of binary variables, \emph{e.g.}, $25$, during each BCD iteration.
This approach  is particularly effective for designing LEO satellite spreading
codes due to the relatively small code lengths and large family sizes.

Finally, penalty methods \cite{YuCYLK2020} and semidefinite relaxations
\cite{DeMaioDHZF2008} have been proposed to design sets of complex-valued,
continuous-phase sequences with constant magnitude. However, we focus on binary
sequences, since they are often preferred in practice due to ease of
implementation. Moreover, the discretization of continuous sequences has been
found to give poor performance \cite{AlaeeMN2019,CuiYFHL2017}.

Other methods construct new sequences and sequence sets by combining
pre-existing binary sequences with desirable correlation properties, such as
Gold codes or optimized sequence sets \cite{BoseS2018,BoukermaRMA2021}.

\section{Spreading code optimization}\label{s-optimization}

\subsection{Preliminaries}

A spreading code family is a set of $m$ binary code sequences, each of length
$n$. We refer to the code family as $X=(x^0,\ldots,x^{m-1})$, where
$x^i\in\{\pm1\}^n$ is the $i$th code in the family. The code family may be
represented as a binary matrix $X\in\{\pm1\}^{n\times m}$, where $x^i$ is the
$i$th column of $X$.

\paragraph{Cross-correlation.} The cross-correlation between two binary codes
$w,v \in\{\pm1\}^n$ is a vector $(w \star v)\in\reals^n$, where
\begin{equation}\label{e-cross-correlation}
(w\star v)_k = \sum_{s=0}^{n-1} w_s v_{(s+k)_{\bmod n}}, \quad k=0,\ldots,n-1.
\end{equation}
That is, $(w \star v)_k$ is the inner product between $w$ and a $k$-circularly
shifted version of $v$. The cross-correlation may also defined for
negative-valued shifts, noting that
\[
(w \star v)_{-k} = (w \star v)_{k-n}, \quad k=0,\ldots,n-1.
\]

\paragraph{Autocorrelation.}
We refer to the cross-correlation of a binary sequence $w\in\{\pm1\}^n$ with
itself as the autocorrelation of $w$. Note that the shift-zero autocorrelation
is given by $(w\star w)_0 = \|w\|_2^2 = n$, regardless of the value of $w$. By
symmetry, $(w\star w)_k = (w\star w)_{-k}$, for all $k=0,\ldots,n-1$.

\paragraph{Autocorrelation sidelobe zero (ACZ).}
A binary sequence $w\in\{\pm1\}^n$ satisfies the ACZ property if its shift-one
autocorrelation $(w\star w)_1$ is minimal. For even-valued $n$, this corresponds
to the requirement that $(w\star w)_1=0$. For odd-valued $n$, the
autocorrelation cannot be zero, and so we instead require that $|(w\star w)_1| =
1$. If the measured correlation at shifts zero and one are too similar, the
receiver may erroneously lose the signal lock. The ACZ property ensures that
the difference between the shift-zero peak and shift-one autocorrelation is
large for all of the codes in the family. This can improve the consistency in
ranging performance across the code family, especially when the code lengths are
short. Indeed, it has been shown that maximizing the difference between the
shift-zero peak and the shift-one autocorrelation minimizes the Cramer-Rao lower
bound for the ranging estimation problem \cite{MedinaOVCVC2020,OrtegaVCC2020}.
The ACZ property has also been applied in practice; the Galileo
constellation uses even-length spreading codes which were designed to satisfy
the ACZ property \cite{Winkel2011,WallnerAHR2007}. In this work, we
formulate the ACZ property as a set of linear inequality or equality
constraints. Therefore, the ACZ property may be readily incorporated into the
MICP formulation of the spreading code optimization problem, as we discuss in
the following subsection.

\subsection{Spreading code optimization problem}

An ideal sequence set $X=(x^0,\ldots,x^{m-1})$ has $(x^i\star x^j)_k$ close to
zero for all pairs of sequences $x^i$ and $x^j$ and at all shifts
$k=0,\ldots,n-1$. In this work, we minimize the sum of squared auto- and
cross-correlation magnitudes, subject to constraint that the autocorrelation
sidelobe zero (ACZ) property is satisfied.

The spreading code optimization problem may be written as
\begin{subequations}\label{e-nonconvex}
\begin{align}
\mbox{minimize}\qquad
& \sum_{i=0}^{m-1}\sum_{j=i}^{m-1}\sum_{k=0}^{n-1}
\left(x^i \star x^j\right)_k^2 & \label{e-nonconvex-obj}\\
\mbox{subject to}\qquad
& \left|(x^i \star x^i)_1\right| \le g,
&i=0,\ldots,m-1, & \label{e-nonconvex-acz}\\
& x^i \in\{\pm1\}^{n}, &i=0,\ldots,m-1. \label{e-nonconvex-binary}
\end{align}
\end{subequations}
Here, $g$ is a parameter that takes value $0$ if $n$ is even, and $1$ if $n$ is
odd. In some contexts, the sum of squares objective is referred to as the
integrated sidelobe level (ISL) \cite{HeLS2012,AlaeeMN2019}. For ease of
notation, we do not exclude the zero-shift autocorrelation terms from the
objective. Since those terms are constant-valued, they do not affect the
solution of the optimization problem.

Note that the objective function \eqref{e-nonconvex-obj} is a non-convex quartic
function of the variables, since each term $\left(x^i \star x^j\right)_k$ is a
nonconvex quadratic function of the binary variables. Similarly, the ACZ
constraint \eqref{e-nonconvex-acz} is a nonconvex constraint. The nonconvex
objective and constraints, combined with the nonconvex binary constraints, make
the problem \eqref{e-nonconvex-obj}--\eqref{e-nonconvex-binary} a challenging
combinatorial optimization problem. In subsection \ref{ss-micp}, we discuss how
the problem may be reformulated as a MICP and how the resulting convex structure
may be exploited \cite{YangMG2023a,YangMG2023b}.

\subsection{MICP formulation}\label{ss-micp}

The cross-correlation function \eqref{e-cross-correlation} involxves a sum of
products of binary variables. By representing each product using a new auxiliary
variable, the cross-correlation becomes a sum of auxiliary variables, and the
objective function \eqref{e-nonconvex-obj} may be written as a convex quadratic
function of the auxiliary variables.

\paragraph{Binary variable product.}
This approach is made possible by the following fact, which may be verified
using a truth table \cite{GloverW1974}. Suppose that $a,b\in\{\pm1\}$ and
$c\in\reals$. Then, $c = ab$ if and only if
\[
\begin{aligned}
c &\le b - a + 1, \\
c &\le a - b + 1, \\
c &\ge -a - b - 1, \\
c &\ge a + b - 1. \\
\end{aligned}
\]
We refer to the above constraints as \emph{linking constraints}, since they
couple the binary variables $a$ and $b$ to the auxiliary variable $c$ that
represents their product.

\paragraph{Spreading code optimization as a MICP.}
Using the aforementioned representation of binary variable products, we may
represent the spreading code optimization problem \eqref{e-nonconvex} as a MICP.
Let $\{z^{i,j}_{s,l}\}$ be a set of auxiliary variables such that
$z^{i,j}_{s,l}$ represents the product $x^i_s x^j_l$, for each
$i,j=0,\ldots,m-1$ and $s,l=0,\ldots,n-1$. Then, the cross-correlation between
codes $x^i$ and $x^j$ may be written as a sum of the auxiliary variables, given
by
\[
(x^i \star x^j)_k = \sum_{s=0}^{n-1} x^i_s x^j_{(s+k)_{\bmod n}}
= \sum_{s=0}^{n-1} z^{i,j}_{s,(s+k)_{\bmod n}}, \quad k=0,\ldots,n-1.
\]
Here, each auxiliary variable $z^{i,j}_{s,k}$ satisfies a set of linking
constraints that couple it to the binary variables $x^i_s$ and $x^j_k$. The
spreading code optimization problem \eqref{e-nonconvex-obj} --
\eqref{e-nonconvex-binary} may therefore be written as
\begin{subequations}\label{e-micp}
\begin{align}
\mbox{minimize}\qquad
& \sum_{i=0}^{m-1}\sum_{j=i}^{m-1} \sum_{k=0}^{n-1}
\left(\sum_{s=0}^{n-1} z_{s,(s+k)_{\bmod n}}^{i,j}\right)^2 & \label{e-micp-obj} \\
\mbox{subject to}\qquad
& -g \le \sum_{s=0}^{n-1} z_{s,(s+1)_{\bmod n}}^{i,i} \le g,
  &i=0,\ldots, m-1, \label{e-micp-acz}\\
& x^i \in\{\pm1\}^{n\times m}, 
  &i=0,\ldots, m-1, \label{e-micp-bin} \\
& z_{s,l}^{i,j} \le x_{l}^j - x_s^i + 1, &\label{e-micp-link1} \\
& z_{s,l}^{i,j} \le x_s^i - x_{l}^j + 1, &\label{e-micp-link2} \\
& z_{s,l}^{i,j} \ge -x_{l}^j - x_s^i - 1, &\label{e-micp-link3} \\
& z_{s,l}^{i,j} \ge x_s^i + x_{l}^j - 1, &\label{e-micp-link4} \\
& i,j=0,\ldots,m-1, \,\, s,l=0,\ldots,n-1. & \nonumber
\end{align}
\end{subequations}
Here, the linear equality constraints \eqref{e-micp-acz} enforces the ACZ
property and the linking constraints \eqref{e-micp-link1} --
\eqref{e-micp-link4} couple the binary variables to the auxiliary variables.
Since the optimization problem \eqref{e-micp} involves minimizing a convex
quadratic function subject to binary, linear inequality, and linear equality
constraints, it is a MICP \cite{ConfortiCZ2014}. More specifically, it is a
mixed-integer quadratic program, since it becomes a quadratic program when the
binary constraints \eqref{e-micp-bin} are relaxed.

\paragraph{Number of auxiliary variables.}
A total of ${nm\choose 2}$ additional auxiliary variables, along with
$4{nm\choose 2}$ linking constraints, are required to transform
\eqref{e-nonconvex} into the MICP \eqref{e-micp}. Therefore, the problem size,
in terms of the number of additional auxiliary variables and constraints, grows
quadratically with $nm$. This may be prohibitive for relevant values of $nm$,
which may be in the tens of thousands. In Section \ref{ss-partial-min}, we show
how the problem may be simplified when we are only interested in optimizing over
only a subset of $B$ of the the binary variables, with the others held fixed. In
that case, the number of auxiliary variables and constraints grows on the order
of $O(B^2)$, rather than $O(n^2m^2)$.

% \paragraph{Alternative objective functions.}
% The approach described in Section \ref{s-bqp} may be used to minimize other
% functions that are convex in the  correlation sidelobe values. Since the auto-
% and cross-correlation functions may be written as linear functions of the
% variables, any convex function of auto- and cross-correlation values is a convex
% function, by the affine composition rule \cite{BoydV2004}. For example,
% replacing the squares of the correlation sidelobes with their absolute values
% (or maximum absolute value) may be formulated as a mixed integer linear program
% (BLP).

\subsection{Partial minimization}\label{ss-partial-min}

In this subsection, we describe the partial minimization of \eqref{e-micp} over
a subset of the binary variables. Note that since \eqref{e-micp} is an MICP, any
partial minimization problem derived from \eqref{e-micp} is also an MICP. The
partial minimization problem is useful for the block coordinate descent
algorithm discussed in Section \ref{ss-bcd}.

Suppose we wish to optimize only over a variable index set 
\begin{equation}\label{e-index-set}
S \subseteq \left\{(i,r) \mid 0\le i \le m-1, \quad 0\le r \le n-1\right\}.
\end{equation}
Each index $(i,r) \in S$ corresponds to the binary variable $x_r^i$, which is
the $r$th element of the $i$th code sequence. The variables not indexed by $S$
are held fixed. That is, we wish to solve the MICP \eqref{e-micp} with the
additional equality constraints
\[
x^i_r = \tilde x^i_r, \quad (i,r)\not\in S,
\]
for some fixed values $\tilde x^i_r \in\{\pm 1\}$, for all $(i,r)\not\in S$.

Since it is only necessary to include auxiliary variables to represent products
between binary variables that appear in $S$, the partial minimization MICP may
be simplified in this case.

In the partial minimization problem, the cross-correlation between any two codes
$x^i$ and $x^j$ may be expressed as an affine function of the auxiliary and
binary variables, with coefficients that depend on the values of the fixed
binary variables. That is, we may write the cross-correlation between two codes
$x^i$ and $x^j$ as
\[
(x^i \star x^j)_k = \sum_{s=0}^{n-1} y^{i,j}_{s,k},
\quad k=0,\ldots,n-1,
\]
where
\begin{equation}\label{e-partial-cross-correlation}
y^{i,j}_{s,k} \vcentcolon= \begin{cases}
z^{i,j}_{s,(s+k)_{\bmod n}}
& \textrm{if $(i,s)\in S$ and $(j,(s+k)_{\bmod n})\in S$}, \\
x^i_s \tilde x^j_{(s+k)_{\bmod n}}
& \textrm{if $(i,s)\in S
\,\,\textrm{and}\,\,(j,(s+k)_{\bmod n}),\,(j,(s-k)_{\bmod n})\not\in S$}, 
\\
\tilde x^i_s \tilde x^j_{(s+k)_{\bmod n}}
& \textrm{otherwise.}
\end{cases}
\end{equation}
There are three cases: both $x^i_s$ and $x^j_{(s+k)_{\bmod n}}$ are variables in
the partial minimization, only one of them is a variable, or neither of them is
a variable.

Since the cross-correlation may be represented using an affine function of the
auxiliary and binary variables, the objective function \eqref{e-micp-obj}
remains convex quadratic. Here, only ${|S| \choose 2}$ auxiliary variables are
required, since we only need to represent products between binary variables
indexed by $S$.

Let $C_S$ be the set of columns, or code sequences, that contain at least one
binary variable indexed by $S$. That is,
\begin{equation}\label{e-column-set}
C_S = \left\{i \mid (i,s) \in S \right\}.
\end{equation}
The number of terms in the objective function \eqref{e-micp-obj} may be reduced
to the auto- and cross-correlation terms between codes in $C_S$. That is, the
partial minimization problem may be written as the MICP
\begin{subequations}\label{e-partial}
\begin{align}
\mbox{minimize}\qquad
& \sum_{i\in C_S}
\sum_{\substack{j=0\\j\ne C_S\,\,\textrm{or}\,\,j\ge i}}^{m-1}
\sum_{k=0}^{n-1}
\left(\sum_{s=0}^{n-1} y_{s,k}^{i,j}\right)^2 & \label{e-partial-obj} \\
\mbox{subject to}\qquad
& -g \le \sum_{s=0}^{n-1} z_{s,(s+1)_{\bmod n}}^{i,i} \le g, \
  &i=0,\ldots,m-1, \label{e-partial-acz}\\
& x^i_s \in\{\pm1\}, &(i,s)\in S, \label{e-partial-bin} \\
& z_{s,l}^{i,j} \le x_{l}^j - x_s^i + 1, \label{e-partial-link1} \\
& z_{s,l}^{i,j} \le x_s^i - x_{l}^j + 1, \label{e-partial-link2} \\
& z_{s,l}^{i,j} \ge -x_{l}^j - x_s^i - 1, \label{e-partial-link3} \\
& z_{s,l}^{i,j} \ge x_s^i + x_{l}^j - 1, \label{e-partial-link4} \\
& (i,s), (j,l) \in S\times S, \nonumber
\end{align}
\end{subequations}
where $y^{i,j}_{s,k}$ are given by \eqref{e-partial-cross-correlation}, and $g$
is a constant that is either $0$ if $n$ is even, or $1$ if $n$ is odd.

\section{Block coordinate descent}\label{s-bcd}

In this section, we describe a block coordinate descent (BCD) method for finding
a good solution to the spreading code optimization problem \eqref{e-micp}. In
our approach, we iteratively solve the partial minimization problem
\eqref{e-partial} over a block, or subset of the binary variables, with the
others held fixed \cite{CuiYFHL2017,YuanLZ2017,YangMG2023a,YangMG2023b}.
The partial minimization problems are solved exactly using an MICP solver, such
as Gurobi \cite{Gurobi} or SCIP \cite{SCIP}. The BCD method is described in
Subsection \ref{ss-bcd}, and variable subset selection strategies are discussed
in Subsection \ref{ss-subset-selection}.

\subsection{Method}\label{ss-bcd}

BCD is a particularly compelling method for handling the MICP \eqref{e-micp},
since the MICP is difficult to solve directly, but its partial minimization
\eqref{e-partial} can be solved effectively in practice.

\paragraph{Basic BCD method.}
The basic BCD method proceeds starting from an initial code family
$X^0\in\{\pm1\}^{n\times m}$. In the $k$th iteration, we compute the next code
family $X^{k+1}$ by performing the following steps:
\begin{enumerate}
\item Select a variable subset $S^k \subseteq \left\{(i,r) \mid 0 \le i \le
m-1,\,0 \le r \le n-1 \right\}$.
\item Solve the partial minimization problem \eqref{e-partial} over $S^k$, with
the other binary variables fixed to their previous values in $X^{k}$.
\item Set $X^{k+1}$ to be the solution to the partial minimization problem.
\end{enumerate}
BCD is a descent method, \emph{i.e.}, the objective value is nonincreasing,
since the block update steps are solved to optimality. We discuss strategies for
selecting the variable subset $S^k$ in Subsection \ref{ss-subset-selection}. The
partial minimization problem \eqref{e-partial} may be solved using an MICP
solver, or exhaustive enumeration.

\paragraph{Two-stage BCD method.} If the code in a given iteration of BCD does
not satisfy the ACZ constraint, then the partial minimization problem
\eqref{e-partial} may be infeasible. That is, it may not be possible to find an
arrangement of the binary variables indexed by $S$ that satisfies the ACZ
constraint. Therefore, we consider a two-stage BCD method, in which the purpose
of the first stage is to find a feasible code family that satisfies the ACZ
constraint. In the first stage, we perform BCD with a modification of the
partial minimization problem \eqref{e-partial}. In the modified partial
minimization problem, the ACZ constraint is removed, and the objective function
is reduced to
\begin{equation}\label{e-jacz}
J=\sum_{i=0}^{m-1}\left(\sum_{s=0}^{n-1} y_{s,1}^{i,i}\right)^2.
\end{equation}
That is, we only minimize the shift-one autocorrelation values.

The first stage is terminated when the ACZ constraint is satisfied. This occurs
when $J=0$ in the even-length case, and when $J=m$ in the odd-length case. We
now discuss a possible termination criterion for the second stage of BCD.

\paragraph{Second-stage stopping criterion.}
When the variable subset size $|S|$ is constrained to be one in every iteration,
the BCD algorithm converges if changing the sign of any single binary variable
does not improve the objective value. In general, if $|S|$ takes a fixed value
$M \ge 1$ in every iteration, then the algorithm converges if changing any $mn
\choose M$ binary variables does not improve the objective value. In practice,
we may terminate the algorithm after the objective value has not improved for a
fixed number of iterations, or after a maximum number of iterations has been
reached.

\paragraph{Initialization.}
The performance of the BCD algorithm depends on the value of the initial code
family. In practice, it may be desirable to run the algorithm multiple times,
initialized with different code families, and select the best solution. The
initial code families may be chosen to be random, or to have desirable
properties. For example, if the initial code family already satisfies the ACZ
property, then the first stage of the two-stage BCD algorithm is unncessary.
Another option is to initialize with a set of codes with good correlation
properties, such as the Gold codes \cite{Gold1967}, Weil codes
\cite{Legendre1808,Rushanan2007} or the output of another optimization method.

\paragraph{Solving MICPs.}
The MICP \eqref{e-micp} and its partial minimization \eqref{e-partial} are both
NP-hard combinatorial optimization problems. In general, those problems can only
be solved by enumerating all possible combinations of binary variables, and the
number of combinations grows exponentially with $nm$. However, the enumeration
may be made more efficient by exploiting the convex structure of the MICP.

In practice, when the variable subsets $S$ are not too large, the partial
minimization problems \eqref{e-partial} may be effectively solved using global
optimization methods such as branch-and-bound \cite{LawlerW1966,BruckerJS1994}
and branch-and-cut \cite{PadbergR1991,StubbsM1999}. The basic idea is that in
each iteration, a convex optimization problem derived from \eqref{e-partial} is
solved, with the binary constraints removed and possibly with additional
variables and convex constraints added. The solution to the convex relaxations
give lower bounds on the optimal value of the original problem, and those lower
bounds may be used to reduce the search space. The commerical solver Gurobi
\cite{Gurobi} and the noncommercial solver SCIP \cite{SCIP} may be used to solve
MICPs.
  
While the aforementioned global methods are often slow and have exponential
worst-case runtime, they can work well when the lower bounds obtained by solving
the convex relaxations are tight. For the partial minimization problems, the
lower bounds are often tight enough to find the global optimum in a reasonable
amount of time, when the number of auxiliary variables required is not too
large.

\subsection{Variable subset selection strategies}\label{ss-subset-selection}

The performance of the BCD algorithm depends on the variable subset selection
strategy. The goal is to choose the variable subset $S$ such that its size can
be made as large as possible, given a computational budget. The time required to
solve the partial minimization problem is the sum of the time required to
compile the MICP \eqref{e-partial} into a form that can be handled by the
solver, and the time required for the solver to find a global solution.

In this work, we select the indices in $S$ randomly in each iteration, with a
sampling scheme that limits the time required to solve the partial minimization
problems in each BCD step.

\paragraph{Limiting the number of active columns.}
The number of active columns in the partial minimization problem is $|C_S|$,
where $C_S$ is given by \eqref{e-column-set}. Since the partial minimization
objective function \eqref{e-partial-obj} involves a sum of with $O(nm|C_S|)$
terms, it may be desirable to limit the number of active columns $|C_S|$,
especially when $n$ and $m$ are large. As seen in Subsection
\ref{ss-subset-sizing}, limiting the number of active columns can greatly reduce
the time required to form and compile the partial minimization MICP in each BCD
iteration. 

\paragraph{Limiting the number of variables in each column.}
Reducing the number of variables in each active column can reduce the time
needed for the MICP solver to find a solution to the partial minimization
problem. As seen in Subsection \ref{ss-subset-sizing}, for a fixed subset size
$|S|$, the time taken by the MICP solver is largest when all of the indices are
concentrated in a single column, and increasing the number of active columns
generally reduces the time taken by the solver. This may be explained by the
fact that the partial minimization problem is more difficult to solve when there
are more variables in each active column. Due to symmetry, an MICP solver based
on branch-and-bound may need to explore a larger number of branches when there
are more variables in each active column \cite{YangMG2023a}.

% \paragraph{Randomizing the variable subset size.}
% In our experiments, we found that mixing BCD steps with large and small variable
% subset sizes can improve the performance of the algorithm. In particular, we
% specify a maximum variable subset size $B$, and select the variable subset size
% uniformly at random from $\{1,\ldots,B\}$ in each iteration.

\section{Results and Discussion}\label{s-results}

In the following, we use the Gurobi optimizer to solve the partial minimization
MICPs involved \cite{Gurobi}. Our implementation has been made publicly
available\footnote{\url{https://github.com/Stanford-NavLab/binary_seq_opt}}.
Subsection \ref{ss-problem-settings} describes the two problem settings
considered in our experiments. Comparisons of variable subset selection
strategies and variable subset sizes are given in Subsections
\ref{ss-subset-experiment}, and \ref{ss-subset-sizing}. 

\subsection{Problem settings}\label{ss-problem-settings}

We considered two problem settings in our experiments, each corresponding to a
different code length $n$ and family size $m$. In each case, we evaluate code
families using the mean-of-squares metric, which is the sum-of-squares objective
\eqref{e-nonconvex-obj}, normalized by the number of terms in the summation.

\paragraph{Set of $66$ length-$127$ codes.}
The first problem setting involves finding a family of $m=66$ binary sequences
each of length $n=127$, and is modeled after the Iridium constellation, which is
a LEO constellation that uses 66 active satellites \cite{Iridium}.

\paragraph{Set of $130$ length-$257$ codes.}
The second problem setting is modeled for potential future LEO PNT
constellations. Xona Space Systems' upcoming LEO constellation is planned to
include 260 satellites \cite{xona2023septentrio}. Since satellites on opposite
sides of the earth (i.e., antipodal satellites) will never simultaneously be in
direct line-of-sight, antipodal satellites can broadcast with the same code
without causing inter-signal interference. Antipodal satellite code sharing
would allow for fewer codes in the complete family, thereby reducing computation
load when the receiver applies correlation processing to search for and acquire
PNT signals. This channel sharing is analogous to the one conducted by GLONASS,
the Russian satellite navigation constellation, which uses frequency division
multiple access~(FDMA) for its G1 signal and assigns antipodal satellites to the
same frequency channel~\cite{MisraE2012}. Therefore, it is sufficient to
consider a families of $m=130$ sequences. In our experiments, we considered
lengths of $n=257$.

\paragraph{Mean-of-squares metric.}
In our experiments, we used the mean-of-squares metric to evaluate code
families. The mean-of-squares metric is defined as the sum of squared
correlation values given by \eqref{e-nonconvex-obj}, normalized by the number of
terms in the summation. That is, the mean-of-squares of a code family
$x\in\{\pm1\}^{n\times m}$ is given by
\begin{equation}\label{e-mos}
J_{\text{MOS}}(x) =
\frac{1}{nm(m+1)/2} \sum_{i=0}^{m-1}\sum_{j=i}^{m-1}\sum_{k=0}^{n-1}
\left(x^i \star x^j\right)_k^2.
\end{equation}

\paragraph{BCD subset sizes.}
For each problem setting, we considered BCD methods with three different
variable subset sizes $|S|$s: $25$, $4$, and $1$. When $|S|=25$, the partial
minimization problem is solved using Gurobi; when $|S|=4$ and $|S|=1$, the
partial minimization problem is solved using exhaustive enumeration. 

\paragraph{Comparison with Gold and Weil codes.}
The BCD methods were compared against the Gold codes \cite{Gold1967} in the case
of $n=127$, and the Weil codes \cite{Legendre1808,Rushanan2007} in the case of
$n=257$. The Gold and Weil codes are well-known families of binary sequences
that are commonly used in satellite communications due to their good correlation
properties \cite{MisraE2012}.

For length $n=127$, there are a total of $129$ Gold codes. Among them, only $65$
satisfy the ACZ constraint. Although there are fewer Gold codes satisfying the
ACZ property than the number of codes in the BCD-optimized code family, the
BCD-optimized codes may still be compared against the Gold codes in terms of the
mean-of-squares objective. For length $n=257$, there are only $128$ Weil codes,
none of which satisfy the ACZ constraint. Although there are fewer Weil codes
than the number of codes in the BCD-optimized code family, the two code families
may also be compared in terms of the mean-of-squares objective.

\subsection{Variable subset selection}\label{ss-subset-experiment}

In this experiment, we compared the time required to solve the partial
minimization problem for the two problem settings, where the variable subset
sizes were fixed to be $|S|=25$ and the number of active columns were varied.
For each active column count $|C_S|$, the number of variable indices in each
active column was limited to be $\lceil 25 / |C_S|\rceil$, \emph{i.e.}, the
variable indices were roughly evenly divided among the active columns. For
example, if we take $|C_S|=4$, then the number of variable indices in each
selected active column is limited to be at most $7$. When $|C_S|=1$, all of the
indices in $S$ were selected from a single column, and when $|C_S|=m$, all of
the indices were selected from different columns.

\begin{figure}
\begin{minipage}[b]{0.95\linewidth}
\centering
\centerline{\includegraphics[width=0.8\textwidth]{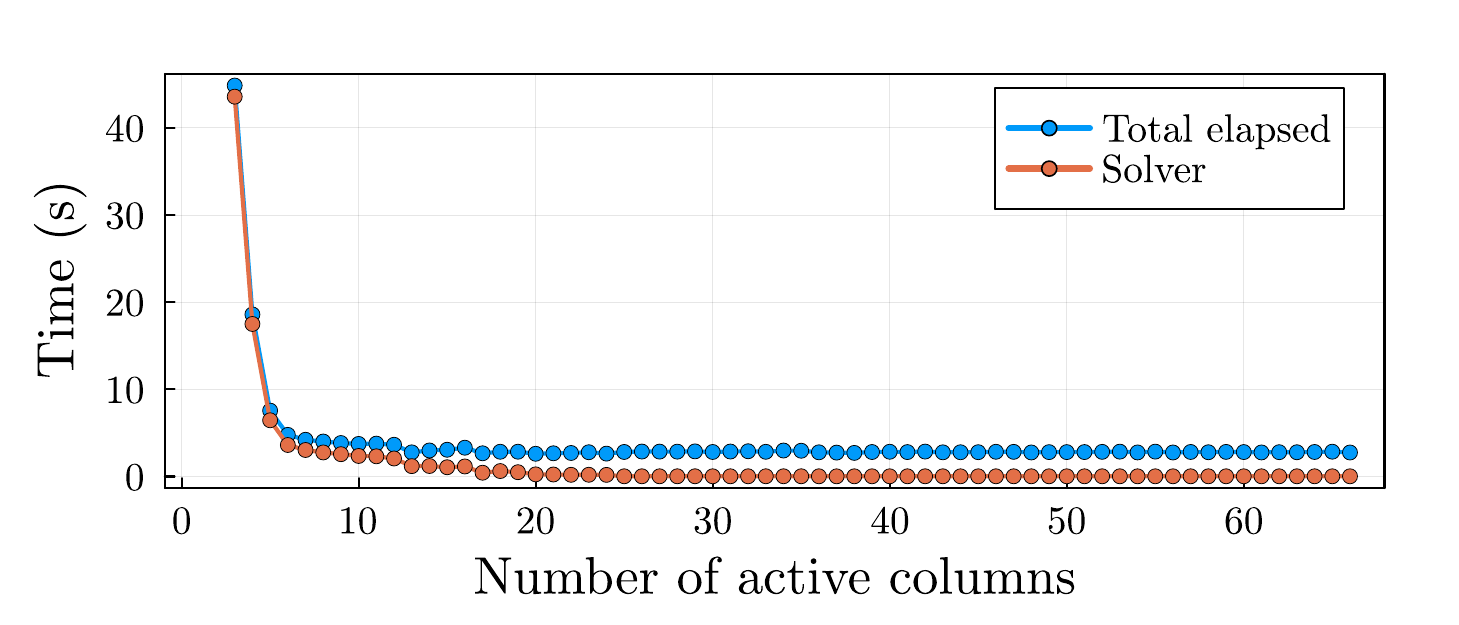}}
\centerline{\includegraphics[width=0.8\textwidth]{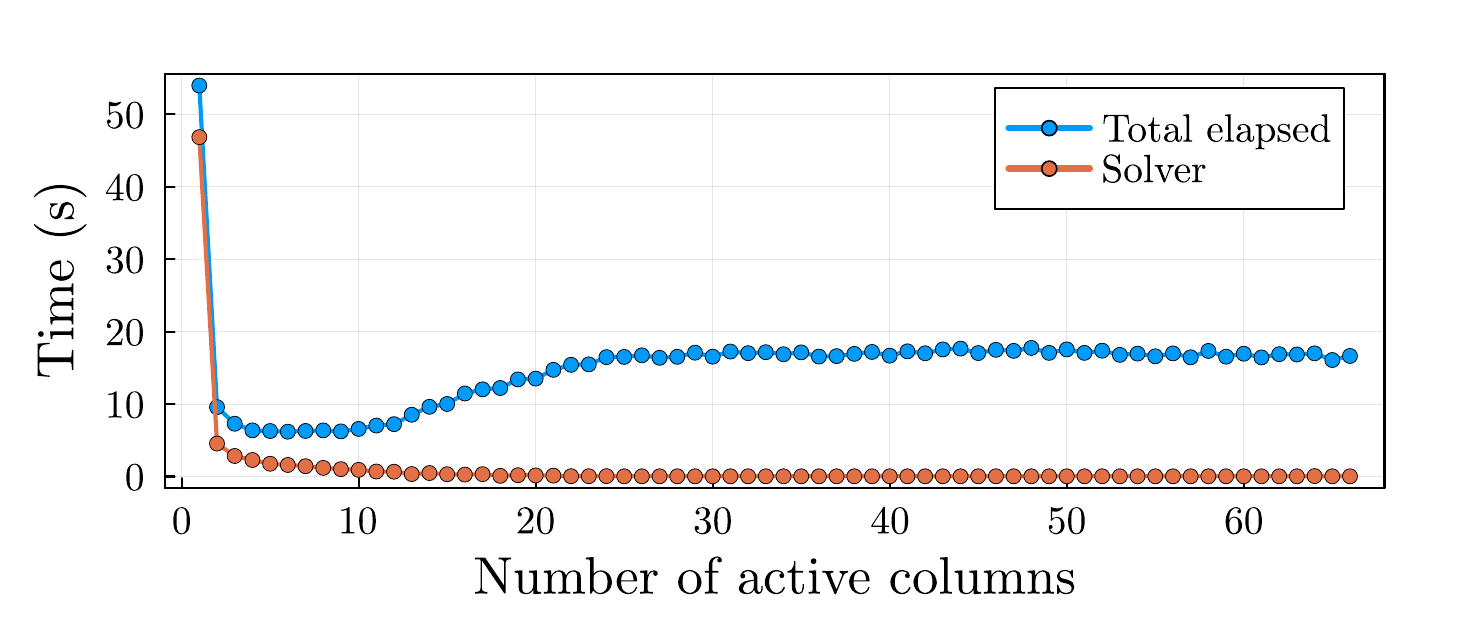}}
\caption{Average BCD iteration time for $n=127$, $m=66$ (top) and $n=257$,
$m=130$ (bottom). In each case, the variable subset size $|S|=25$, and the
number of indices in each active column is limited to $\lceil|S|/|C_S|\rceil$
(\emph{i.e.}, the indices were roughly evenly divided among the active columns).
Each plotted point is the average of $30$ runs, where each run was computed
using a random code family and random indices $S$. We show both the total time
taken to form and solve the partial minimization problems, as well as the time
taken by only the solver (Gurobi).}%\medskip
\label{fig:bcd_timing}
\end{minipage}
\end{figure}

Figure \ref{fig:bcd_timing} compares the average time needed to solve the
partial minimization problem for different numbers of active columns. The time
taken by the Gurobi MICP solver is plotted, along with the total elapsed time,
which also includes the time required to form and compile the partial
minimization MICPs. Each plotted point is the average time taken over $30$ runs,
where each run involved a random code family and a random variable subset $S$.
The random code families were generated uniformly at random, and the partial
minimization problems were solved without the ACZ constraint.

For the $n=127$, $m=66$ problem instance, the amount of time required to solve
the partial minimization problem decreases monotonically with the number of
active columns. However this is not the case in the $n=257$, $m=130$ case. In
that case, the total time required initially decreases with the number of active
columns, but then increases again. This is due to the overhead required to
compile the MICP into a form that can be handled by the solver, since the time
taken by the solver itself decreases monotonically with the number of active
columns.

% \subsection{Fixed versus variable subset sizes}

% In this experiment, we compared the performance of the BCD algorithm when the
% variable subset sizes were fixed to be $|S|=25$ in each iteration, and when the
% variable subset sizes were selected uniformly at random from $\{1,\ldots,25\}$.
% Based on Figure \ref{fig:bcd_timing}, we chose the indices in $S$ to all be from
% different columns for the $n=127$ case, \emph{i.e.}, $|C_S|=1$ in each
% iteration. In the $n=257$ case, we chose $S$ such that five indices were
% selected from five different columns, \emph{i.e.}, $|C_S|=5$ in each iteration.

\subsection{Comparison of BCD methods}\label{ss-subset-sizing}

\begin{figure}
\begin{minipage}[b]{0.95\linewidth}
\centering
\includegraphics[width=0.49\textwidth]{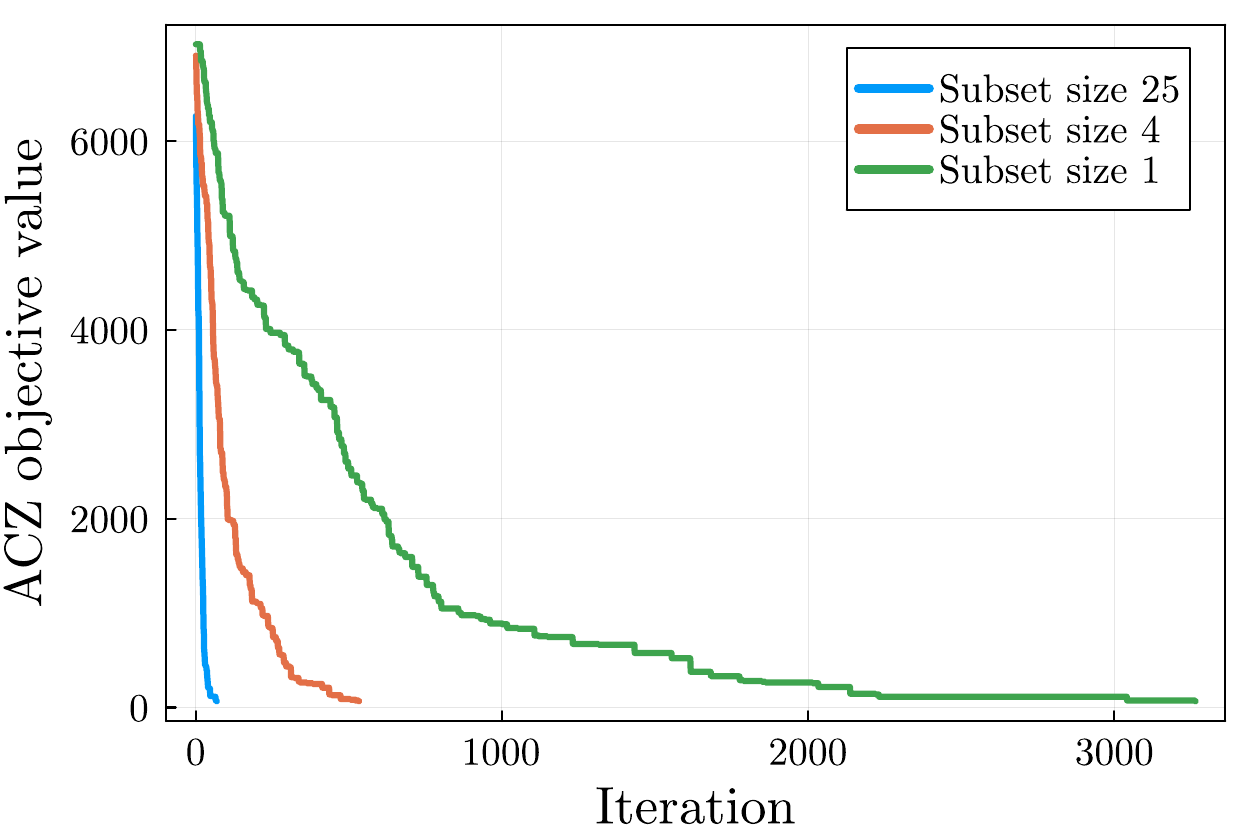}
\includegraphics[width=0.49\textwidth]{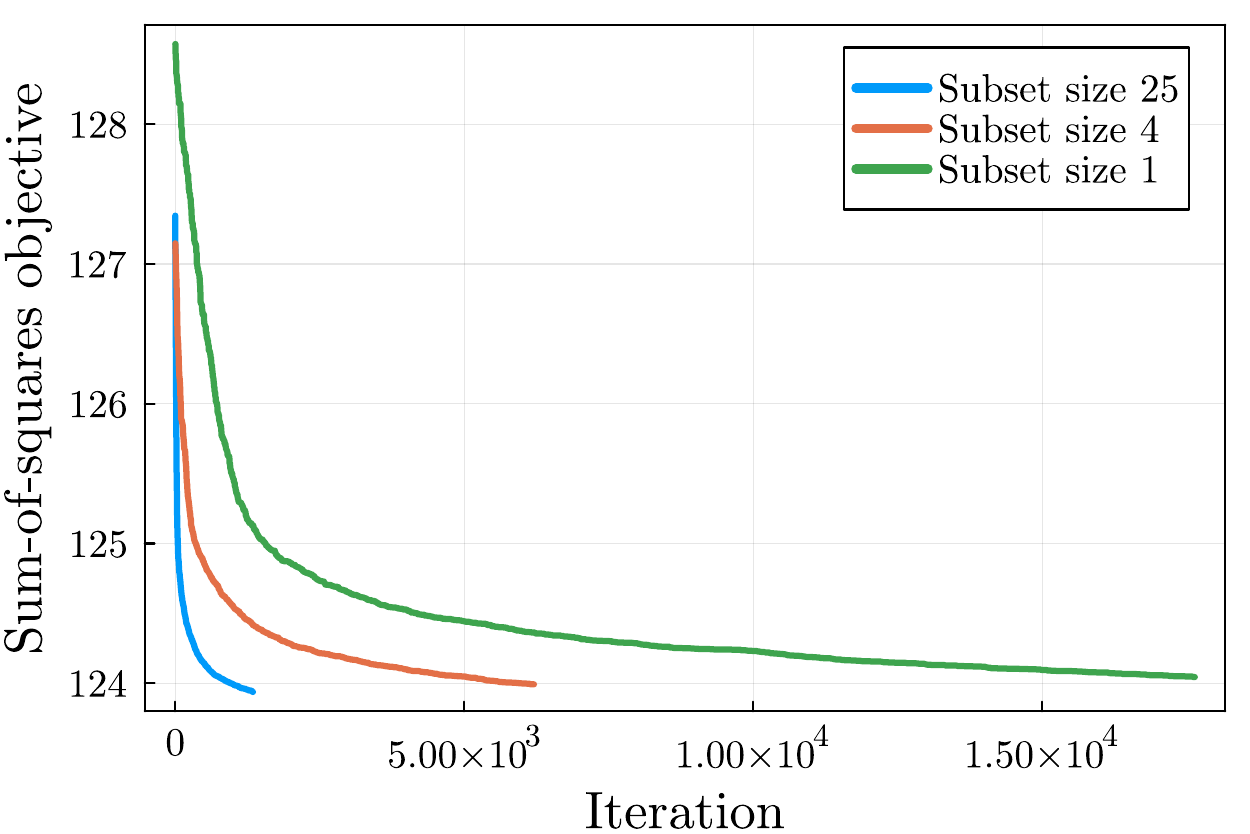}\\
\includegraphics[width=0.49\textwidth]{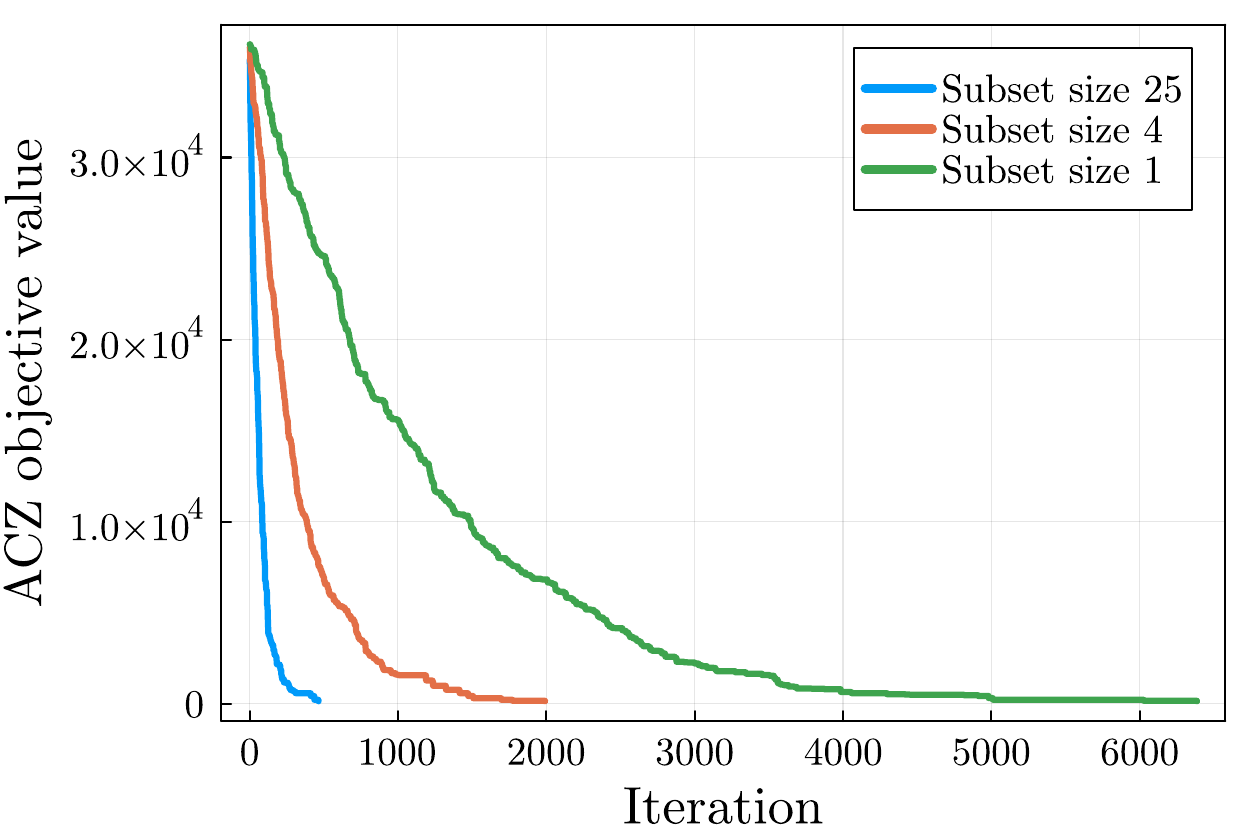}
\includegraphics[width=0.49\textwidth]{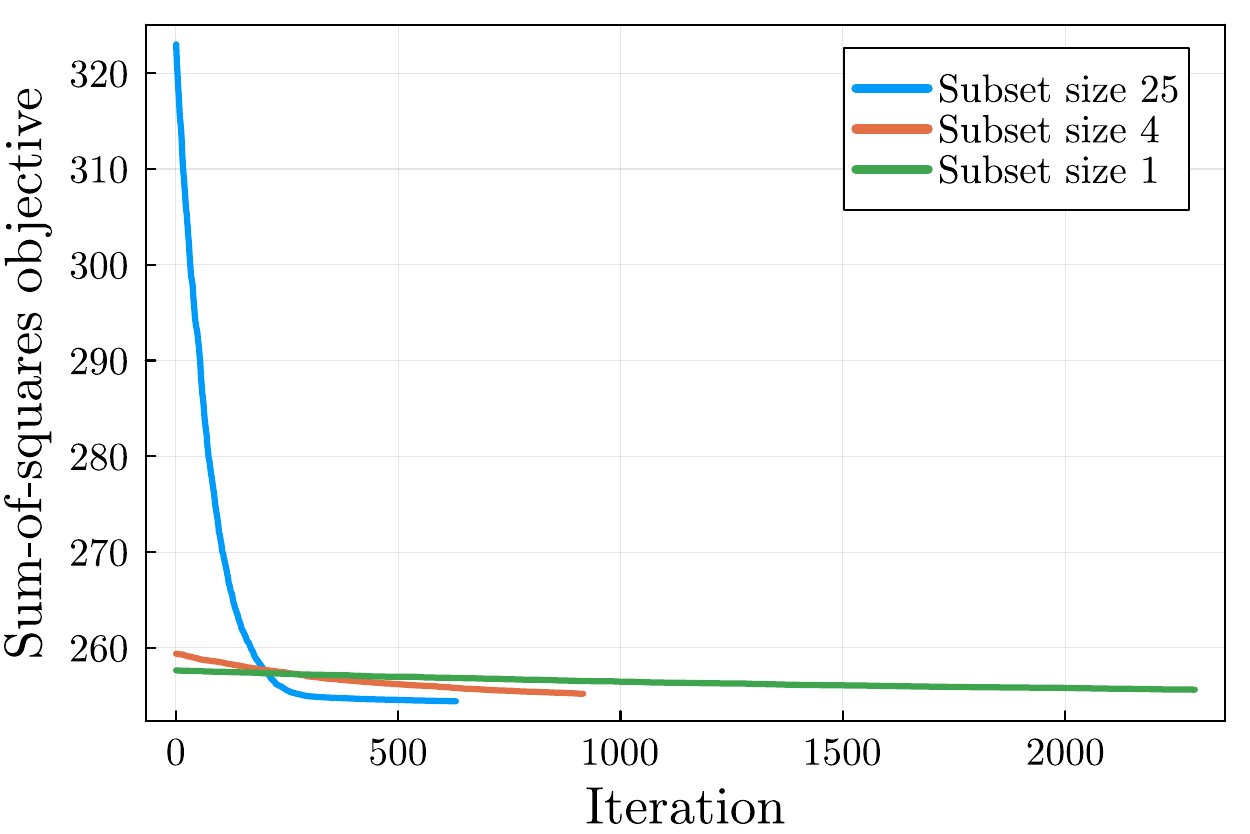}
\caption{Top: objective value vs. iteration for stage-one (left) and stage-two
(right) BCD, for problem setting $n=127$. Bottom: objective value vs. iteration
for stage-one (left) and stage-two (right) BCD, for problem setting $n=257$.
Stage-one BCD (left) is terminated when the ACZ property is satisfied. Objective
values for stage-two BCD (right) are shown for the first hour of computation.
}%\medskip
\label{fig:bcd_performance}
\end{minipage}
\end{figure}
  
Next, we evaluate the performance of the two-stage BCD method with variable
subset sizes $|S|=25$, $|S|=4$, and $|S|=1$.

\paragraph{Variable subset selection strategy.}
We use the following selection scheme, which is based on the results in
Subsection \ref{ss-subset-experiment}. In the $n=127$ case, the variable indices
were selected by choosing a single variable index from $|S|$ different columns.
In the $n=257$ case, the variable indices were selected by choosing the indices
at random, with the constraint that the number of the number of active columns
and the number of variable indices in each active column were limited to five
each.

\paragraph{Results.}
Figure \ref{fig:bcd_performance} shows the objective value vs. iteration for the
three subset sizes. The top left and bottom left plots show the ACZ objective
value \eqref{e-jacz} vs. iteration for the first stage of BCD, for the $n=127$
and $n=257$ problem settings, respectively. The first stage is terminated when
the ACZ constraint is satisfied. The top right and bottom right plots show the
mean-of-squares metric \eqref{e-mos} vs. iteration for the second stage of BCD,
for the $n=127$ and the $n=257$ problem settings, respectively. The plots show
the objective values for the first hour of computation.

In each case, it can be seen that increasing the subset size $|S|$ leads to a
lower objective in fewer iterations, but also fewer total iterations, since the
cost of each iteration is higher. Table \ref{table:gold_weil} compares the
mean-of-squares of the BCD-optimized codes with the Gold and Weil codes, where
the BCD methods were run for $12$ hours. The BCD-optimized code with $|S|=25$
found a code with the lowest mean-of-squares in both problem settings. Table
\ref{table:iteration-counts} shows the number of BCD iterations taken in the
$12$ hour period, for each subset size $|S|$.

\begin{figure}
\begin{minipage}[b]{0.95\linewidth}
\centering
\includegraphics[width=0.49\textwidth]{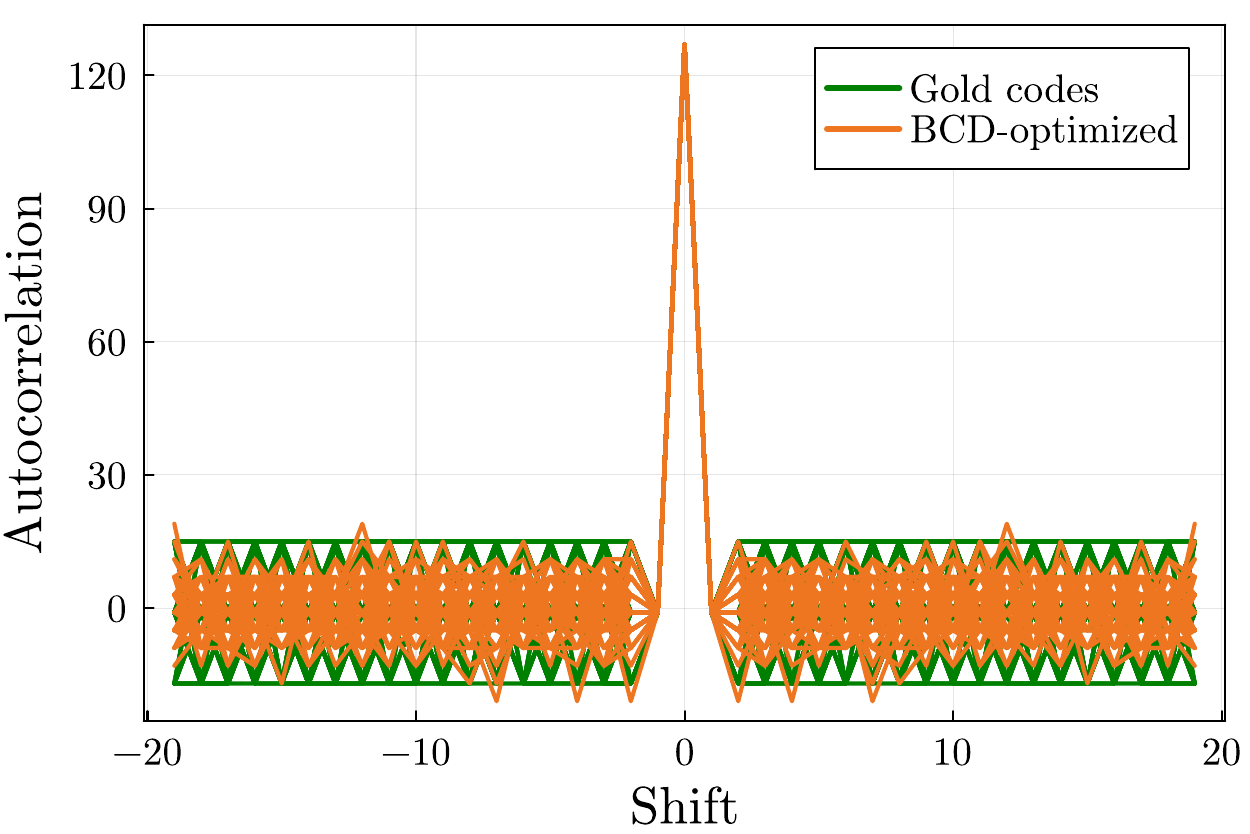}
\includegraphics[width=0.49\textwidth]{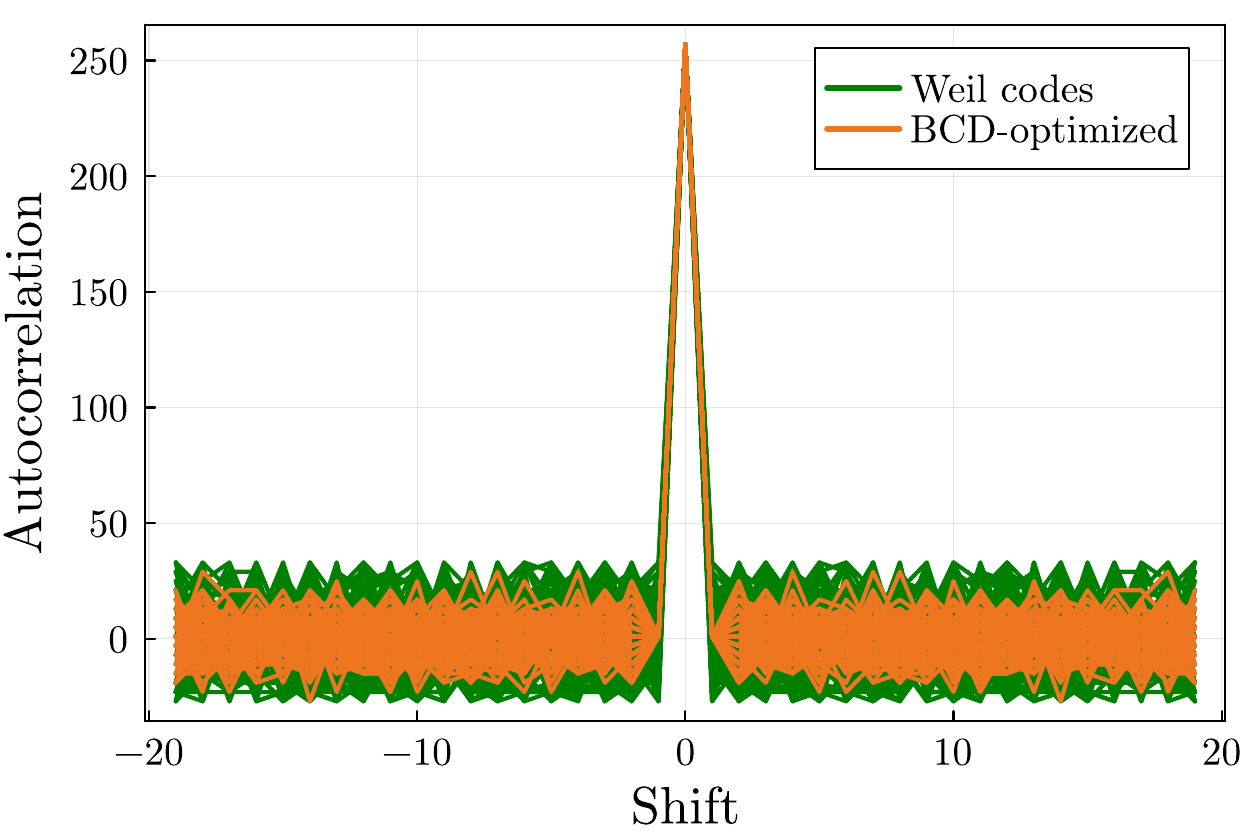}
\caption{Autocorrelation of one of the 66 optimized binary sequence, compared
with the autocorrelation of one of the Gold codes.}%\medskip
\label{fig:bcd_autocorr}
\end{minipage}
\end{figure}

\paragraph{Autocorrelation visualization.}
Finally, Figure \ref{fig:bcd_autocorr} shows a superposition of the
autocorrelations of the codes found using the BCD method with subset size $25$,
compared with a superposition of the autocorrelations of the Gold and Weil
codes. Both the BCD-optimized codes and the selected Gold codes satisfy the ACZ
constraint, while it may be seen that the Weil codes do not. The BCD-optimized
codes appear to strictly outperform the Weil codes. While the BCD-optimized
codes appear to have autocorrelation closer to zero than the Gold codes on
average, they have a larger peak autocorrelation magnitude than the Gold codes.

\begin{table}
\centering
\captionsetup{width=0.95\textwidth} % Adjust the width as you like
\caption{Comparison of mean-of-squares of BCD optimized codes with Gold and Weil
codes. In the $n=127$ case, BCD was used to optimize over $m=66$ codes, while
there were only $m=65$ corresponding Gold codes. In the $n=257$ case, BCD was
used to optimize over $m=130$ codes, whereas there were only $m=128$ Weil codes.
The BCD-optimized and Gold codes satisfy the ACZ constraint, while the Weil
codes do not.}
\label{table:gold_weil}
\begin{tabular}{ccc}
& $n=127$ & $n=257$ \\
\midrule
BCD ($|S|=25$) & \textbf{123.741} & \textbf{253.707} \\
BCD ($|S|=4$) & 123.745 & 254.092 \\
BCD ($|S|=1$) & 123.762 & 254.234 \\
Gold & 125.95 & - \\
Weil & -  & 255.99 \\
\end{tabular}
\end{table}

\begin{table}
\centering
\captionsetup{width=0.95\textwidth} % Adjust the width as you like
\caption{Number of BCD iterations taken in a 12-hour period, for different subset sizes $|S|$.}
\label{table:iteration-counts}
\begin{tabular}{ccc}
& $n=127$ & $n=257$\\ \hline
$|S|=25$ & 15,756 & 8,680 \\
$|S|=4$ & 70,966 & 10,278 \\
$|S|=1$ & 209,966 & 26,671 \\
\end{tabular}
\end{table}

\section{Conclusions}\label{s-conclusion}

In this work, we considered the problem of designing binary spreading codes with
good auto- and cross-correlation properties, in particular for LEO applications,
where the number of codes are large, relative to the code lengths. We
proposed a two-stage BCD method for finding codes of arbitrary length and family
size that both satisfy the ACZ property and have good correlation properties.
We demonstrated that the method can find codes that outperform the Gold and Weil
codes in terms of the mean of squared correlation values.

Finally, we discuss possible directions for future work. First, the BCD
method proposed in this work may be extended to account for the effects of
Doppler shift, which can be significant in LEO navigation settings
\cite{Soualle2005}. For example, the BCD method may be used to optimize the
average of the objective function \eqref{e-nonconvex-obj} over a range of
Doppler shifts \cite{YangMG2024}. Second, this work did not consider the
effects of any data or secondary codes, which are often superimposed on the
primary spreading codes \cite{WallnerAHR2007}. Those superimposed codes may
adversely affect the correlation properties of the primary codes, and may be
worth consideration in future work.

\backmatter

\section*{Declarations}%% if any

\subsection*{Ethics approval and consent to participate}
Not applicable.

\subsection*{Consent for publication}
Not applicable.

\subsection*{Availability of data and materials}
The datasets generated and/or analysed during the current study are available in
the \texttt{binary\_seq\_opt} repository,
\url{https://github.com/Stanford-NavLab/binary_seq_opt}

\subsection*{Competing interests}
The authors declare that they have no competing interests.

\subsection*{Funding}
This material is based upon work supported by the Air Force Research Lab (AFRL)
under grant number FA9453-20-1-0002.

\subsection*{Authors' contributions}
AY formulated an extension of the MICP formulation from prior work, using the ASZ property, and implemented and ran the experiments. 
TM assisted in the experiment setup with the computing cluster, design of the experiments for the LEO navigation setting and code design objectives. 
GG assisted in ideation of the problem context for LEO navigation.  
All authors read and approved the final manuscript.

\subsection*{Acknowledgments}
Not applicable.

\bibliography{sn-bibliography}% common bib file
%% if required, the content of .bbl file can be included here once bbl is generated
%%\input sn-article.bbl

\end{document}